\documentclass[aps,prl,reprint,twocolumn,superscriptaddress]{revtex4-1}
\usepackage{amsmath}
\usepackage{amssymb}
\usepackage{graphicx}

\begin{document}

\title{Two-dimensional double-quantum spectra reveal collective resonances in an atomic vapor}

\author{Xingcan Dai}

\altaffiliation[Present address: ]{Department of Physics, Tsinghua University, Beijing 100084 China.}

\affiliation{
    JILA, University of Colorado and National Institute of Standards and Technology, Boulder, Colorado 80309-0440
}

\author{Marten Richter}

\altaffiliation[Present address: ]{Institut f\"{u}r Theoretische Physik, Nichtlineare Optik und Quantenelektronik, Technische Universit\"{a}t Berlin, Hardenbergstr. 36, 10623 Berlin, Germany.}

\affiliation{
    Department of Chemistry, University of California, Irvine, California 92697-2025
}

\author{Hebin Li}
\affiliation{
    JILA, University of Colorado and National Institute of Standards and Technology, Boulder, Colorado 80309-0440
}

\author{Alan D. Bristow}

\altaffiliation[Present address: ]{Department of Physics, West Virginia University, Morgantown, West Virginia 26506-6315 USA.}

\affiliation{
    JILA, University of Colorado and National Institute of Standards and Technology, Boulder, Colorado 80309-0440
}

\author{Cyril Falvo}

\altaffiliation[Present address: ]{Institut des Sciences Mol\'{e}culaires d Orsay, UMR CNRS 8214, Universit\'{e} Paris Sud 11, 91405 Orsay, France.}

\affiliation{
    Department of Chemistry, University of California, Irvine, California 92697-2025
}

\author{Shaul Mukamel}
\affiliation{
    Department of Chemistry, University of California, Irvine, California 92697-2025
}

\author{Steven T. Cundiff}
\affiliation{
    JILA, University of Colorado and National Institute of Standards and Technology, Boulder, Colorado 80309-0440
}

\date{\today}

\begin{abstract}
We report the observation of double-quantum coherence signals in a gas of potassium atoms at twice the frequency of the one-quantum coherences. Since a single atom does not have a state at the corresponding energy, this observation must be attributed to a collective resonance involving multiple atoms. These resonances are induced by weak inter-atomic dipole-dipole interactions, which means that the atoms cannot be treated in isolation, even at a low density of $10^{12}$ $\mathrm{cm}^{-3}$.
\end{abstract}

\pacs{34.20.Cf, 82.50.Pt, 82.53.Kp}
%\keywords{}

%comment out for word count
\maketitle

%\section{}
%\subsection{}
%\subsubsection{}

\ Resonant interactions occur in a wide range of physical and chemical systems, from nuclear spin-spin coupling \cite{Ernstbook} to excitonic effects in light harvesting antennae and reaction centers \cite{Engel2007,Collini2010}. Two-dimensional Fourier transform (2DFT) methods excel in isolating and characterizing these interactions.
Multidimensional Fourier transform techniques were originally developed in nuclear magnetic resonance, where they proved very powerful in disentangling congested spectra and providing structural information of complex molecules by measuring the strength of inter-nuclear couplings \cite{Ernstbook}. Thanks to recent developments in ultrashort pulse laser technology, these techniques have been extended into the infrared and visible regions of the electromagnetic spectrum during the last two decades \cite{Mukamel2000,Cho2008}. When visible light is used, the method is often called 2D electronic spectroscopy because electronic transitions are probed. Measurements have provided evidence that quantum coherence plays a role in photosynthesis \cite{Engel2007,Collini2010} and given insight into many-body interactions in semiconductor nanostructures \cite{Zhang2007}. The ability of 2DFT signals to dissect the quantum pathways of matter contributing to the nonlinear response makes it a powerful spectroscopic tool.

\ Quantum pathways that include a double-quantum coherence can be observed in 2DFT spectroscopy for the appropriate time ordering of the excitation pulses. They are particularly interesting because theoretically they have been shown to be highly sensitive to the presence of many-body interactions \cite{Yang2008} and can reveal fine details of many-body wavefunctions. Double-quantum coherences have been observed in 2DFT spectra using infrared excitation of molecular vibrations \cite{Fulmer2004,Sul2006}, in semiconductor quantum wells \cite{Stone2009, Karaiskaj2010} and in molecular electronic transitions \cite{Christensson2010,Nemeth2010}. These signals characterize coherences between the ground state and a doubly excited state, which is often taken as proof that a doubly excited state exists in the level structure of an individual chromophore. However, we show that collective resonances resulting from inter-atomic interactions can induce double-quantum coherences. It has been suggested that such resonances can be induced in non-interacting atoms by entangled photons \cite{Muthukrishnan2004,Richter2011}.
Interaction-induced double-quantum coherences may play a role in photo-processes such as light harvesting \cite{Engel2007,Collini2010}. Furthermore, quantum information processing schemes have been proposed that exploit collective resonances due to long-range dipole-dipole coupling \cite{Lukin2001}.

\ Here, we report the observation of double-quantum coherences in a gas of potassium atoms. Based on the level scheme of a single potassium atom, shown in Fig. 1(a), and the spectrum of the excitation laser, the na\"{i}ve expectation is that double-quantum coherences should not exist because there are no atomic levels at twice the frequency of the laser, which is tuned to be resonant with the one-quantum transitions corresponding to the $D_1$ and $D_2$ lines. We interpret the observed two-quantum resonances as being due to collective excitations of two atoms, and use a Hilbert space transformation to provide a heuristic explanation of how interactions result in an effective level at twice the one-quantum energy. For simplicity, we first discuss two-level atoms, as the essential physics is not changed because we excite both $D_1$ and $D_2$ transitions.  A more rigorous theoretical treatment using an excitonic picture and incorporating dipole-dipole interactions in a gas of atoms combined with molecular dynamics simulations reproduces the experimental results.

\begin{figure}
\includegraphics[width=.8\columnwidth]{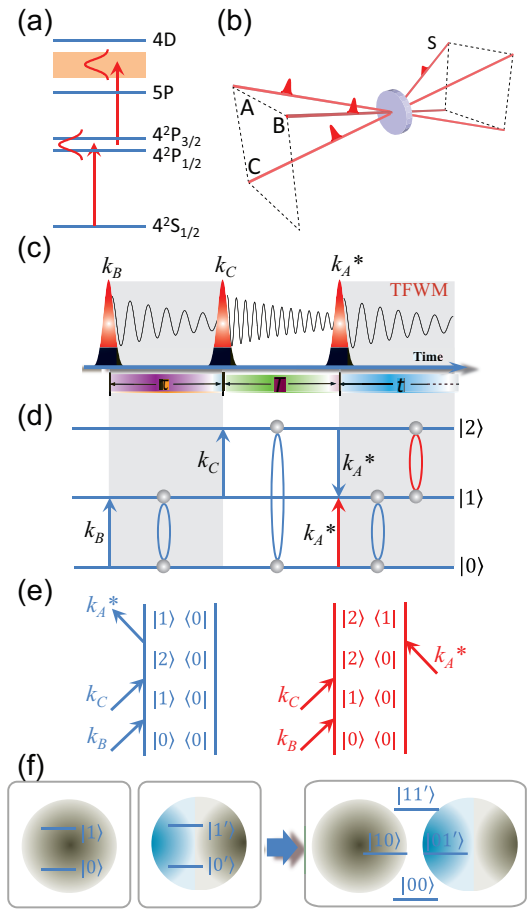}
\caption{\label{figure1}(color online) (a) Levels of an isolated potassium atom; (b) Geometry of incident beams and signal beam; (c) Pulses and oscillating coherences during different time periods; (d) Transitions of a ladder-level system as a function of time showing transitions driven by laser pulses and resulting coherences; (e) Double sided Feynman diagrams for the two quantum pathways that contribute to $S_{III}$ for a ladder level scheme; (f) Hilbert space transformation between two independent two-level atoms and a 4-level system including a doubly excited state.}
\end{figure}

\ The quantum pathways that include a double-quantum coherence are distinguished by the combination of the wave-vector and time-ordering selection of the pulses. As shown in Figure 1(b), the three incident pulses, with wavevectors $\mathbf{k}_A$, $\mathbf{k}_B$ and $\mathbf{k}_C$, focus and intersect at the sample such that they are on three corners of a square that is perpendicular to the propagation direction. Their interaction produces a signal in the direction $\mathbf{k}_S = -\mathbf{k}_A + \mathbf{k}_B+ \mathbf{k}_C$, which corresponds to the 4th corner of the square after the sample. Double-quantum coherences can contribute to the signal only if the pulse with the conjugated wavevector, $\mathbf{k}_A$, arrives last. This excitation sequence is shown schematically in Fig. 1(c) and (d) for a simplified 3-level system consisting of a ground state, $|0\rangle$, singly excited state, $|1\rangle$, and double excited state, $|2\rangle$. The first pulse, $\mathbf{k}_B$, puts the system in a quantum mechanical superposition between states $|0\rangle$ and $|1\rangle$, i.e., in a single quantum coherence. The second pulse converts the single quantum coherence to a double quantum coherence between $|0\rangle$ and $|2\rangle$. The third pulse converts the double quantum coherence back to a single quantum coherence that radiates. The final single-quantum coherence can be between states $|0\rangle$ and $|1\rangle$ or between states $|1\rangle$ and $|2\rangle$, both of these possibilities are shown in Fig 1(d). These pathways can be described using the two double-side Feynman diagrams for the atomic density matrix shown in Fig. 1(e).

\ Experimental two-dimensional single- and double- quantum spectra are acquired using the apparatus described in Ref. \onlinecite{Bristow2009}. A mode-locked Ti:sapphire laser generates $\sim200$ fs pulses that are input to an ultrastable platform of nested and phase stabilized interferometers to generate 4 identical pulses arranged in a box geometry. Three of the beams are the excitation beams, $\mathbf{k}_A$, $\mathbf{k}_B$ and $\mathbf{k}_C$, while the fourth is designated the tracer and propagates in the same direction as the signal beam, $\mathbf{k}_C$. The tracer is used to generate a reference pulse that is routed around the sample and interfered with the emitted signal beam to produce interferograms. The full phase and amplitude information about the signal can be extracted from the interferograms. The tracer itself is blocked during data acquisition. It is used to determine the overall phase of the signal for decomposition into real and imaginary parts \cite{Bristow2008}.

The potassium vapor is held in a 350 micron thick transmission cell. The cell body is made of titanium with two sapphire windows that are diffusion bonded to the titanium. The vapor temperature is controlled by a heater attached to the cell. For the measurement reported here, the cell temperature was 130 C. The transmitted spectrum of the strongly attenuated laser was used to estimate the absorbance. The transmitted intensity is $I_t = I_i e^{-\alpha l}$, where $I_i$ is the incident intensity, $\alpha$ is the absorption coefficient and $l$ is the cell thickness. At this temperature and density, the resonance broadened linewidths are smaller than the Doppler width and spectrometer resolution, thus a 1550 torr Argon buffer gas was introduced into the cell to induce collision broadening and reduce the peak absorbance $\alpha l$ to 0.053 for the $D_2$ line, which is stronger than the $D_1$ line. This low absorbance rules out optical density effects as explaining the observations.

Figure 2 shows the real part of both single- and double-quantum spectra for a potassium number density of $3.5\times 10^{12}$ cm$^{-3}$. As observed in previous measurements \cite{Dai2010}, the single-quantum spectrum shows peaks corresponding to the $D_1$ and $D_2$ lines, as well as off-diagonal peaks due to coupling between them via the ground-state bleach and Raman-like coherences. Surprisingly, the double-quantum spectrum also shows clear peaks even though there are no atomic states at these energies. The observed linewidths correspond to a dephasing time slightly longer 5 ps for the double-quantum coherences. The $D_1$ and $D_2$ lines correspond to transitions from the $4^2S_{1/2}$ ground state to the $4^2P_{1/2}$ and $4^2P_{3/2}$ states, with transition frequencies of 389.29 and 391.02 THz, respectively. The higher lying states that are closest to twice the $D_{1,2}$ energies are the $5P$ and $4D$ states, which are at frequencies 740.81 and 821.36 THz, both of which are well outside the spectral range of shown in Fig. 2(a) and well outside the laser bandwidth of 3 THz. The fact that the observed resonances are at exactly twice the frequencies of the $D_1$ and $D_2$ lines (or their sum frequency) indicate that the double-quantum resonances are due to the combined response of two atoms, rather than the level structure of a single atom and that the coupling is weak (otherwise  the resonance would be shifted).

\begin{figure}
\includegraphics[width=.8\columnwidth]{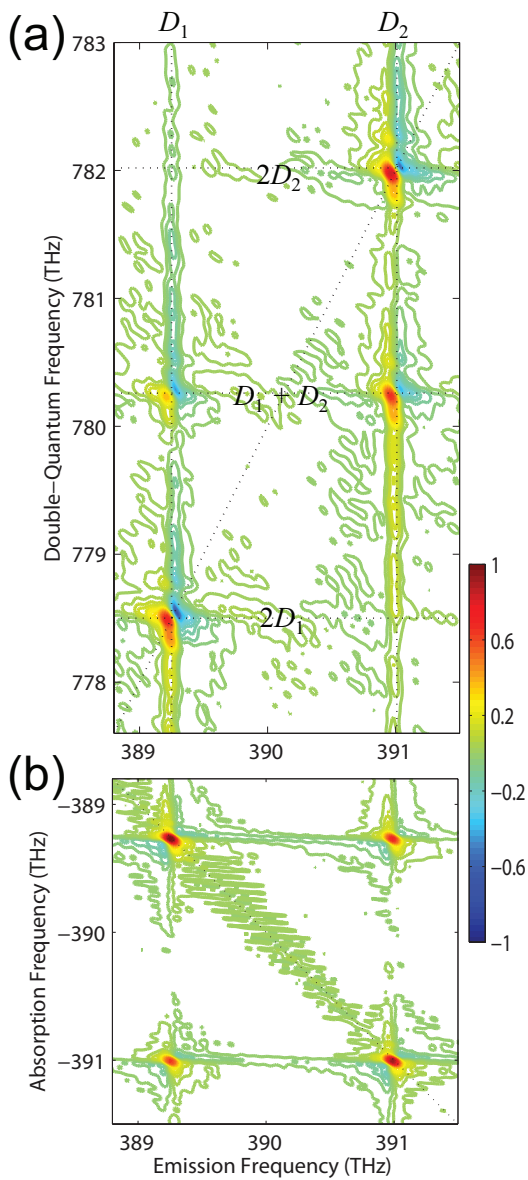}
\caption{\label{figure2} (color) Experimental (a) double-quantum and (b) single-quantum real spectra.}
\end{figure}

\ The observations can be explained using a simple picture obtained by combining the Hamiltonians of two individual atoms into a Hamiltonian describing both simultaneously, as shown in Fig. 1(f), and working in the joint Hilbert space of both atoms. For two isolated atoms, each described as a two level system, there are no doubly-excited levels, so a double-quantum coherence is impossible. However, the joint space of the two two-level atoms has 4-levels. These consist of a ground state, where both atoms are in the individual ground states, two singly excited states where one of the two atoms is excited, and a doubly excited state where both of them are excited. In this four-level system, double-quantum coherences can occur between the ground state and the doubly excited state. However, this change of description should not affect the underlying physics, which seems contrary to the fact that the nonlinear susceptibility calculated for the four-level system has terms that are not present for two isolated two-level systems. Careful analysis shows that all of these ``new'' terms for the four-level system interfere destructively and   exactly cancel if the ground state to singly-excited state transitions are identical to the singly-excited to doubly-excited state transitions. For the double-quantum coherences, this means that the two pathways shown in Fig. 1(e) cancel one another. Thus describing the system in the joint Hilbert space alone does not change the nonlinear susceptibility. The picture changes if there are interactions between the atoms. In that case, the transitions will not be identical and the cancelation will be incomplete, resulting in double-quantum resonances. The joint Hilbert space description is useful in explaining why interactions result in double-quantum resonances. This picture only considers the interaction of a pair of atoms; a more complete theoretical treatment needs to consider an interacting ensemble.

\ To go beyond this simple explanation, we performed simulations using the exciton formalism. Excitons are delocalized collective electronic excitations of assemblies of coupled chromophores including coherences that reflect entanglement. Excitons with extended coherence size were first discussed in crystals \cite{Frenkel1931,Wannier1937}. In real crystals, disorder and dynamic fluctuations limit the coherence size. An atomic vapor is an ideal model system for studying excitonic effects in the presence of disorder \cite{2008Lorenz_PRL}. In atomic vapors, $n$-exciton states correspond to states where $n$ atoms are excited and all others are in the ground state.  The atoms interact due to dipole-dipole coupling.

The single- and double-quantum spectra are simulated based on the Nonlinear Exciton Equations (NEE) \cite{Mukamel2000}. The effective Frenkel exciton Hamiltonian, which describes the electronic excitation of interacting atoms, is
\begin{equation}
\begin{split}
H(t)&=\sum_{m}\Omega_{m}(t)\hat{B}_{m}^{\dagger}\hat{B}_{m}+\\
&\sum_{m,n,m\neq
n}(J_{mn}(t)\hat{B}_m^\dagger\hat{B}_n+K_{mn}(t)\hat{B}_m^\dagger\hat{B}
_n^\dagger\hat{B}_m\hat{B}_n)
\end{split}
\end{equation}
where $\hat{B}_{m}^{\dagger}$ and $\hat{B}_{m}$ are the exciton creation and annihilation operators on state $m$, $m$ is a multi-index consisting of the atom number and the intra-atom exciton state(e.g. states connected to $D_1$ or $D_2$ line). $\Omega_{m}$ is the excitation energy of exciton $m$, $J_{mn}$ indicates the one-exciton coupling strength between excitons $m$ and $n$ and $K_{mn}$  is the two-exciton coupling parameter. The dipole-dipole coupling model for $J_{mn}$ can be calculated from

\begin{equation}
J_{mn}(t)=\dfrac{1}{4\pi\varepsilon_0\varepsilon}(\dfrac{\vec{\mu}_m\cdot\vec{
\mu}_n}{R_{mn}(t)^3}-\dfrac{3(\vec{\mu}_m\cdot\vec{R}_{mn}(t))(\vec{\mu}
_n\cdot\vec{R}_{mn}(t))}{R_{mn}(t)^5})
\end{equation}
	
where  $\vec{\mu}_{m,n}$ are the transition dipole moments and  $R_{mn}$ is the distance between atoms $m$ and $n$, $\varepsilon$ is the dielectric constant. The Hamiltonian for the coupling of the atoms to the optical field is
\begin{equation}
\begin{split}
H_{int}(t)&=-E(\vec{r},t)\cdot V(t)\\
&=-E(\vec{r},t)\cdot\sum_{m}\vec{\mu}_m(t)(\hat{B}_{m}^{\dagger}(t)+\hat{B}_{m}
(t)).
\end{split}
\end{equation}
All elements of the Hamiltonian are time dependent through the changing configuration of the potassium atoms.
%Thus the four-wave-mixing signal that is proportional to the third order response function can be calculated.

The simulation of the double-quantum coherence spectra is based on a direct nonlinear exciton propagation method in an exciton picture \cite{Falvo2009}. The simulation of the 2DFT spectra requires three steps: first, the trajectories of an ensemble of atoms are simulated; second, the trajectories of the atoms are used to calculate the time-dependent interatomic couplings for atoms; third, the 2DFT spectra are generated using the calculated interatomic couplings. Dynamic trajectories of 20 potassium atoms in a simulation unit cell were generated using the molecular dynamics package GROMACS. The   ground state potential energy surface of $\mathrm{K}_2$ is used as the coupling function between atoms. Runs up to 2 ns were performed with a resolution of 0.01 ps. Then the trajectories of atoms are used to calculate the interatomic distances between atom pairs and the coupling parameter $J_{mn}$ for each snapshot of the trajectory using Eq.~2. The intra-atomic elements of $K_{mn}$ are set to large numbers to ensure that each atom cannot be excited twice, while the interatomic elements of  $K_{mn}$ are set to zero since the interatomic two-exciton couplings are ignored. As the last step, 2D spectra are generated using the SPECTRON package \cite{Falvo2009}. The double-quantum coherence spectrum, $S_{III}(\tau,T,t)$, is
\begin{eqnarray}
 &&S_{III}(\tau,T,t)=2 \left(\frac\imath\hbar\right)^4
\sum_{t_s} \sum_{n_4} \vec{\mu}_{n_4}(\tau+T+t+t_s) \nonumber\\
&&\qquad \times R^{S_{III}}_{n_4}(\tau+T+t+t_s,
T+t+t_s,t+t_s,t_s).
\end{eqnarray}
The signal is averaged over several start times, $t_s$, along the trajectory with
\begin{eqnarray}
R^{S_{III}}_{n_4}(\tau_4,\tau_3,\tau_2,\tau_1)&=&\sum_{m_4 m_1}\int_{\tau_3}^{\tau_4}\mathrm{d}s
G_{n_4,m_4}(\tau_4,s) K_{m_4 m_1}(s)  \nonumber\\
&&\qquad
\psi^{(2)}_{m_4 m_1}(s;\tau_2;\tau_1)\psi^{(1)*}_{m_1}(s;\tau_3)
\end{eqnarray}
for the one-exciton Green function $G_{n_2,n_1}(\tau_2,\tau_1)=\langle g| B_{n_2} U(\tau_2,\tau_1)B^\dagger_{n_2}|g\rangle$ where $U(\tau_2,\tau_1)$ is the time-evolution operator connected to $H(t)$ . The excitonic functions are computed by direct integration of the Schr\"odinger equation $\imath\hbar \partial_t |\psi^{(1)}(t;\tau_3)\rangle=H(t)|\psi^{(1)}(t;\tau_3)\rangle$
and $\imath\hbar \partial_t |\psi^{(2)}(t;\tau_3)\rangle=H(t)|\psi^{(2)}(t;\tau_3)\rangle$  with the initial conditions $\psi^{(1)}_{m_1}(t=\tau_1;\tau_1)=\vec{\mu}_{m_1}(\tau_1)\cdot E(\tau_1)$
and
\begin{eqnarray}
\psi^{(2)}_{m_1,m_2}&&(t=\tau_2;\tau_2;\tau_1)= \nonumber\\ &&\frac{1}{\sqrt{2}}\{\vec{\mu}_{m_2}(\tau_2)\cdot E(\tau_2) \psi^{(1)}_{m_1}(t=\tau_2;\tau_1) \nonumber\\
&&+ \vec{\mu}_{m_1}(\tau_2)\cdot E(\tau_2) \psi^{(1)}_{m_2}(t=\tau_2;\tau_1)\}. \nonumber
\end{eqnarray}

The double-quantum spectrum shown in Fig. 3(a) is obtained by Fourier transforming $S_{III}(\tau,T,t)$ with respect to $\tau$ and $t$. The single-quantum spectrum (Fig. 3(b)) was calculated similarly as for the double-quantum spectrum. The model parameters have been adjusted to match the experiment; however our goal is not a detailed fit to the experimental results, but rather to simply show that double-quantum resonances do appear. The simulations do show double-quantum coherences, which do not appear if the dipole-dipole coupling is turned off.  Furthermore, the simulated lineshapes are in good agreement with the experiment, for example both display a dispersive profile, which has been shown to be characteristic of many-body interactions in semiconductors \cite{Li2006}.

\begin{figure}
\includegraphics[width=.8\columnwidth]{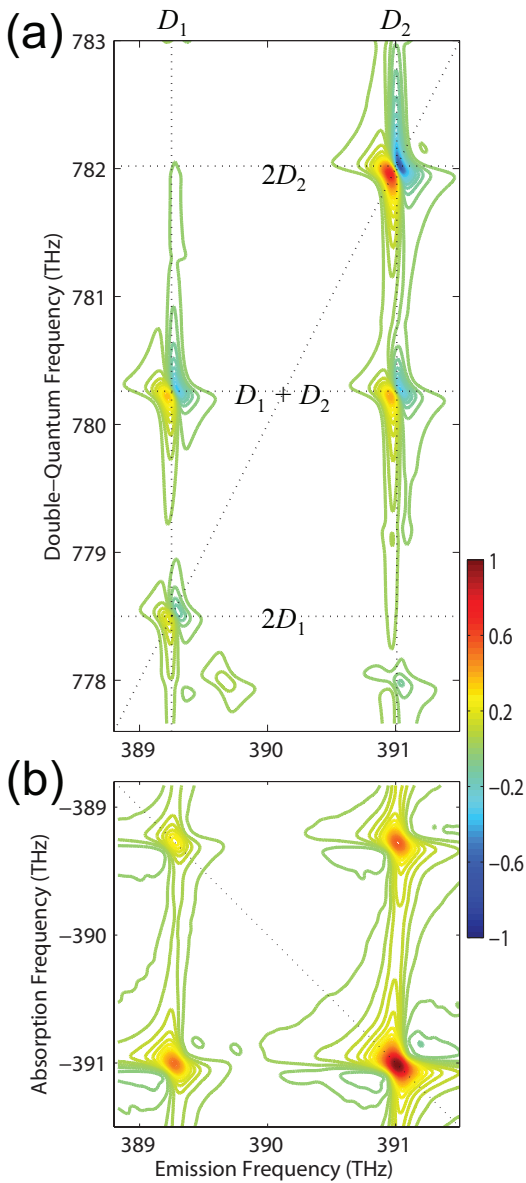}
\caption{\label{figure3} (color) Simulated (a) double-quantum and (b) single-quantum real spectra.}
\end{figure}

\ Our results demonstrate that, thanks to its high sensitivity, electronic double-quantum coherence spectroscopy can detect even very weak interactions between chromophores, which makes it a powerful technique, but at the same time means that the interpretation of double-quantum spectra may not be possible based on the level structure of a single chromophore and thus collective effects should be considered. Our results are for the regime where any spectral shifts are small compared to linewidth and laser bandwidth, as compared to observations fo a Rydberg blockade that were in the opposite regime \cite{Farooqi2003}. Moreover, the long-range nature of the dipole-dipole interaction means that this effect is present even at low densities, where it is often thought that interactions can be neglected. Collective coupling to the radiation field is also responsible for enhanced spontaneous emission or superradiance \cite{Ariunbold2010}. We find that the ratio of the strength of the double-quantum coherences to the single-quantum coherences is constant as the atomic density is varied over an order of magnitude. This result is consistent with the fact that self-broadening is theoretically found to be independent of density \cite{Leegwater1994} due to the $r^{-3}$ scaling of the dipole-dipole coupling.

%If you have acknowledgments, this puts in the proper section head.
\begin{acknowledgments}
We thank T. Asnicar and H. Green for technical assistance. The work at JILA was supported by NIST and the NSF Physics Frontier Center program. S.T.C. is a staff member in the NIST Quantum Physics Division. The work at UC-Irvine was supported by the NSF Chemistry Division and DARPA. M.R. acknowledges support from the Alexander von Humboldt Foundation through the Feodor Lynen Program.
\end{acknowledgments}

%\bibliography{KvaporS3}


\end{document}